\documentstyle[12pt]{ioplppt}

\newcommand{\beq}{\begin{equation}}  
\newcommand{\eeq}{\end{equation}}  
\newcommand{\bea}{\begin{eqnarray}}  
\newcommand{\eea}{\end{eqnarray}}  

\begin{document}
\title{Echoes in classical dynamical systems}

\author{Bruno Eckhardt}

\address{Fachbereich Physik, Philipps-Universit\"at Marburg,
35032 Marburg, Germany}

\begin{abstract}
Echoes arise when external manipulations
to a system induce a reversal of its time evolution
that leads to a more or less perfect recovery of the 
initial state. We discuss the accuracy with which a
cloud of trajectories returns to the initial
state in classical dynamical systems that
are exposed to additive noise and small differences
in the equations of motion for forward and
backward evolution. 
The cases of integrable and chaotic motion and
small or large noise are studied in some detail and
many different dynamical laws are identified.
Experimental tests in 2-d flows that show chaotic
advection are proposed.
\end{abstract}
%
%
\maketitle

\section {Introduction}

Echoes arise when through suitable manipulations in a system
the dynamics is reversed and a more or less complete 
recovery of the initial state is achieved.
Acoustical echoes arise from 
reflections of sound at walls, 
spin echoes from reversals of magnetic fields 
(Hahn 1950, Carr and Purcell 1954),
current echoes through a sequence of
suitable electromagnetic pulses (Niggemeier \etal 1993)
 and 
Loschmidt echoes from a reversal of momenta in a 
Hamiltonian system (Loschmidt 1876). 
That echoes can also appear in many particle systems
is at first surprising since it seems to 
be in conflict with the irreversibility implied by the second
law of thermodynamics. Closer inspection shows, however,
that the recovery of the initial state is not perfect,
and studies of the deviations tell a lot about the 
mechanisms that break reversibility.

Several aspects of echo phenomena in dynamical systems have
recently been studied in connection with Loschmidt echoes
in quantum systems (triggered by Pastawski \etal 1995
and Levstein \etal 1998). 
An initial state $|0\rangle$ is propagated forward in time with
Hamiltonian $H$ and then back in time with a slightly
different Hamiltonian $H'$ (as suggested 
by Peres 1984).  The loss of coherence is
measured in terms of the fidelity 
$\langle 0 | e^{iH't/\hbar} e^{-iHt/\hbar} | 0\rangle$.
This is the same as calculating the overlap between the 
states $|t\rangle =  e^{-iHt/\hbar} | 0\rangle$ and
$|t'\rangle =  e^{-iHt'/\hbar} | 0\rangle$, obtained by
propagating the same initial state $| 0 \rangle$ for the
same time $t$ but with two different Hamiltonians. The
decay of the overlap as a function of time and difference
in Hamiltonian and the various time regimes have been the
subject of several recent papers, e.g.
(Jalabert and Pastawski 2001,
Jacquod, Silvestrov and Beenakker 2001, 
Tomsovic and Cerruti 2002, 
Benenti and Casati 2002,
Wisniacki and Cohen 2001,
Prosen and Znidaric 2002,
Prosen and Seligman 2002).

The present paper is devoted to Loschmidt echoes in classical dynamical systems.
If there is no difference between forward and backward equations of motion,
the initial state is recovered perfectly. But what happens if there are small
differences or if the system is exposed to noise? And what are the differences
between echo experiments in integrable and chaotic systems? These questions will
be addressed for Gaussian densities
in linearized flows: they provide a convenient and sufficiently general 
class of densities in which a large variety of dynamical behaviour can be identified.

Besides the obvious connection to the quantum echo experiments,
the calculations are of some relevance for two other,
directly classical situations:
numerical trajectory reversals and reversibilty in advection.

When the equation 
of motion are reversed in numerical calculations trajectories 
will typically not return to their starting point. For chaotic systems,
the inherent sensitivity to initial conditions suggest an exponentially large
deviation. A lack of growth has been used as an indicator for quantum
regularity (Casati et al 1986). Obviously,
no such problems should arise for perfect reversals and
perfect numerical integrators since the solutions to
the equations of motion are uniquely specified by the
initial conditions (baring singular points in the
differential equations). The fact that trajectories
do not return to their starting points thus reflects
numerical inaccuracies from finite time steps and
limited resolution. Using Gaussians to represent a cloud of initial
conditions, noise to reflect truncation errors and differences between
forward and backward integration algorithm this numerical
reversibility experiment can be connected to the problem considered
here.

A popular and impressive demonstration of echoes in classical
systems is provided by flow reversals in viscous liquids:
a blob of dye can be stretched out until it is barely visible
but upon reversal of the flow it reforms almost completely!
The multimedia fluid mechanics CD (Homsey et al 2001) contains
several demonstrations in laminar flows. The connection to chaos
comes through experiments on chaotic advection 
(Aref 1984, 2002, Ottino 1990). 
For instance, in their experiments
on chaotic advection, Chaiken \etal (1986) noted
that if the dye passed through a chaotic region the recovery
was less perfect, but they did not investigate this in 
detail. Since many of the typical flows can be realized experimentally,
also the dynamics of the Gaussians discussed below should be 
experimentally accessible.

The types of system considered here are classical.
They may be exposed to additive white noise and
forward and backward motion may differ.
The state of the system is characterized
by a smooth density in phase space, and the evolution equation
is the Fokker-Planck equation with appropriate drift term.
The discussion will be limited to 
densities that are Gaussian in shape, for which a  
reasonably complete general discussion is possible.
Superpositions of such Gaussians can be used
to approximate other densities.
General expressions in arbitrary dimensions will be given,
but their implications are most evident in 2-d conservative flows:
this is also the case that is accessible in hydrodynamical
systems.

The outline of the paper is as follows. In the next section
the dynamics of the center of mass and the variances
for Gaussian densities will be discussed. In section 3 this 
information is applied to 
the discussion of echoes in integrable systems
(section 3a) and chaotic ones (3b). Section 4 contains
a discussion of the results and some remarks on 
experimental tests in 2-d advection systems.

\section{Gaussian densities}
\subsection{Outline of echo experiments}
The calculation of an echo naturally divides
into two steps: the forward evolution up to some
time $T$ under one set of equations,
followed by the backward evolution under a perhaps
slightly different set of equations for the 
same time interval. If a mapping of initial conditions
under a class of time evolutions can be found, say
$\rho_{f} = U_1(T) \rho_i$ for the forward
evolution under flow ${\bf u}_1$, then we can write for 
the backward evolution with flow ${\bf u}_2$ the formal
expression
$\rho_{b} = U_2(-T) \rho_{f}$, so that the mapping
to the echo state $\rho_e$ becomes
\beq
\rho_e = U_2(-T) U_1(T) \rho_i\,.
\eeq
Thus, on the technical level it suffices to find
$U$ for the forward dynamics of a sufficiently
large class of densities and systems. Note that in the 
presence of noise or in a dissipative system there is a 
difference between (i) comparing the initial state with the state
obtained by propagating an initial condition over the complete
cycle of forward and backward evolution, and (ii) 
comparing the states obtained by propagating the same initial condition 
forward in time with two different flows: dissipation and noise
break the reversibility that permitted the change in 
protocol in the quantum case.

The evolution equation for the densities ${\rho}({\bf x},t)$ 
is the Fokker-Planck equation,
\beq
\dot\rho = - \nabla({\bf u} \rho) + D \Delta \rho
\eeq
where $D$ is the molecular diffusion constant.
Echoes are most easily identified when initial and final density
are sufficiently similar, as is the case for strongly localized
objects. More complicated initial densities may be approximated 
by superpositions of localized ones. The dynamics for localized
densities splits, in leading order in moments, into two parts, 
the motion of the center of
mass and the changes in shape and size, as measured by the
variances. This expansion may be extended to higher order
moments of the density, but the equations become
too cumbersome to analyze.

\subsection{Center of mass motion}
The center of mass of a localized density follows a 
classical trajectory ${\bf x}_P(t)$, where
\beq
\dot {\bf x}_P(t) = {\bf u}({\bf x}_P(t),t)\,.
\eeq
The notation here is borrowed from hydrodynamic advection
(Aref 1984, 2002),
where ${\bf u}$ is a velocity field and ${\bf x}_P$ the
trajectory of a particle advected by the fluid. In other
situations the velocity field ${ \bf u } ( { \bf x },t )$ has 
to be replaced by the right hand side ${\bf f}$ of an evolution
equation $\dot{\bf x}={\bf f}({\bf x},t)$. 

For the backward propagation, where a small modification of the 
flow field is permitted, the trajectory may differ from the one
during forward evolution and we need to estimate the differences
between the two. 
Let ${\bf x}_P(t)$ be a trajectory in a velocity field ${\bf u}$
and ${\bf x}_P(t)+{\bf q}(t)$ one in the 
perturbed velocity field ${\bf u}+\delta{\bf u}$.
To first order in ${\bf q}$ the equation becomes
\beq
\dot{\bf q} (t) = A(t) {\bf q}(t) + \delta{\bf u}({\bf x}_P(t),t)\,,
\eeq
where $A$ is the linearization of the full velocity
field ${\bf u}+\delta{\bf u}$ at the trajectory
${\bf x}_P$,
\beq
A_{ij} = \frac{\partial (u_i+\delta u_i)}{\partial x_j}({\bf x}_P(t),t) \,.
\label{A_def}
\eeq
With the help of the monodromy matrix $M(t)$, the solution to
\beq
\dot M = A M
\eeq
with initial condition $M(0)=1$, a formal solution can be
given,
\beq
{\bf q}(t) = M(t) \left( q(0) + \int_0^t d\tau M^{-1}(\tau)
\delta{\bf u}({\bf x}_P(\tau),\tau)\right)\,.
\label{displace}
\eeq
This is the general solution for the displacement in 
a perturbed velocity field. A discussion of specific
examples will be deferred to section 3 below.

\subsection{Variances}
With $\tilde{\bf x}={\bf x}-{\bf x}_P(t)$ the coordinates relative
to the center of mass, the density can be written as 
$\rho(\tilde{\bf x},t)=\rho({\bf x}-{\bf x}_P(t), t)$ 
and the Fokker-Planck equation becomes
\beq
\dot\rho(\tilde{\bf x},t) = 
- \nabla({\bf u}({\bf x}_P+\tilde{\bf x},t) \rho) +
({\bf u}({\bf x}_P,t)\cdot\nabla)\rho + D \Delta \rho\,,
\eeq
where now all spatial derivatives are with respect to 
the relative coordinate $\tilde {\bf x}$.
The localization of the densities allows a linearization
of the velocity field near the trajectory, i.e.
\beq
{\bf u}({\bf x}, t) = {\bf u}({\bf x}_P(t),t) + A(t) \tilde{\bf x}
\eeq
with the derivative matrix $A$, eq.~(\ref{A_def}).
Then
\beq
\dot\rho = - (\tr A) \rho - ( (\tilde{\bf x}^T A^T) \cdot\nabla) \rho 
+ D \Delta \rho\,.
\eeq
For conservative Hamiltonian systems and incompressible
flows, the trace of $A$ vanishes; the discussion will
henceforth be limited to that case.
The next step in the analysis is to note that the equations
are second order with a linear position dependence at most,
so that a solution in terms of Gaussian densities is possible
(Eckhardt 1990).
In an $n$-dimensional phase space with variance matrix $\Gamma(t)$
they are given by
\beq
\rho(\tilde{\bf x},t) = \pi^{-n/2} \left(\det \Gamma\right)^{-1/2}
e^{-\tilde{\bf x}^T \Gamma^{-1} \tilde {\bf x}}
\label{rhoi}
\eeq
where $\Gamma$ satisfies
\beq
\dot \Gamma = 2 D + A \Gamma + \Gamma A^T
\label{Gamma_eq}
\eeq
(note that $\Gamma$ as a kernel for a quadratic form is symmetric).
With the help of the monodromy matrix $M(t)$ 
also a closed expression for the variance matrix can be given,
\beq
\Gamma(t) = M(t) \left(
{\Gamma_i + 2D\int_0^t \left( M(\tau)^T M(\tau)\right)^{-1} d\tau }\right)
M(t)^T \,.
\label{gt}
\eeq
This is the central formula for the dynamics of the variances
on which the calculation of the various situations can be based.
For the specific cases studied in the next section a direct
solution of (\ref{Gamma_eq}) was found to be simpler.

\section{Echoes}
\subsection{Classical fidelity}
The analysis of echoes now proceeds as follows: We start with an
initial density $\rho_i$ with variance matrix $\Gamma_i$.
This density evolves under the influence of a velocity field
${\bf u}$ and additive white noise
for some time $T$. At the end of this
time interval the position of the 
density is ${\bf x}_P(T)$ and the variance is
$\Gamma_f(T)$, as given by (\ref{gt}).
Then the field is reversed. If the reversal is perfect, 
the new velocity field ${\bf u}'$ equals $-{\bf u}$ and the 
center of mass returns exactly. If there is a slight deviation
the center of mass will move along a trajectory with a small
displacement ${\bf q}(t)$ according to (\ref{displace}).
In both cases, however, the variance does not return exactly, unless 
all noise is suppressed, i.e. $D=0$. 
This holds quite generally, the only approximation being
the Gaussian shape of the density and 
linerization of the velocity field near the trajectory.

A general Gaussian echo may then be displaced
with its center of mass by ${\bf q}$ and will have
a variance matrix $\Gamma_e$. For a comparison between initial and 
echo density of states the positions and variances
can be used directly. But it is also possible to 
mimick the quantum fidelity expression and to introduce
an overlap between classical phase space densities. The main 
difference is that quantum wave functions are normalized within
the $L^2$ norm and classical densities are not. The proper
definition of the fidelity as the cosine of the angle
between the two densities in Hilbert space then is
\bea
{\cal O} &=& \frac{\int d{\bf x} \rho_i({\bf x}) \rho_e({\bf x})}{
\left(\int d{\bf x} \rho_i({\bf x})^2\right)^{1/2}
\left(\int d{\bf x} \rho_e({\bf x})^2\right)^{1/2}
}\,.
\eea
Without the normalization the overlap could change even if
both densities evolve in the same way. The definition given 
by Prosen and Znidaric (2002) thus has to be modified.
For the case of two Gaussians, an initial density $\rho_i$ 
centered at zero and with
variances $\Gamma_i$, and an echo density
\beq
\rho_e = \pi^{-n/2} \left(\det \Gamma_e\right)^{-1/2}
e^{-({\bf x}-{\bf q})^T \Gamma_e^{-1} ({\bf x}-{\bf q})}
\eeq
the overlap becomes
\bea
{\cal O} &=& 2^{n/2} \sqrt{
 \frac{\sqrt{\det \Gamma_i} \sqrt{\det \Gamma_e}}%
{\det(\Gamma_i+\Gamma_e)}}
e^{-{\bf q}^T \left(\Gamma_i^{-1}+\Gamma_e^{-1}\right) {\bf q}}\,.
\eea

The overlap integral indicates two
very different kinds of contributions: The prefactor
measures the reduction in overlap by spreading of the density.
The exponential factor accounts for the rapid
drop off in overlap when the centers of the Gaussians are separated;
the Gaussian form is clearly connected to the Gaussian 
tails in the density, and would be different, e.g., for 
exponential tails in the density.

\subsection{Shear flows}
In laminar flows neighboring trajectories see 
slightly different velocities and separate
linearly in time. When combined with noise
a cubic growth of the variance results.
Specifically, consider a 2-d shear flow
\beq
{\bf u} = \left(\matrix{\alpha y\cr 0}\right)
\eeq
of shear rate $\alpha$. The associated monodromy matrix is
\beq
M(t) = \left(\matrix{1& \alpha t\cr 0 & 1} \right)\,.
\eeq
During forward evolution the variances become
\bea
\Gamma_{11}^{(f)} &=&  \Gamma_{11}^{(i)} + 2\alpha \Gamma_{12}^{(i)} T
+  \alpha^2 \Gamma_{22}^{(i)} T^2 + 2DT + \frac{2}{3}\alpha^2 DT^3
\label{g111}\\
\Gamma_{12}^{(f)} &=&  \Gamma_{12}^{(i)} + \alpha \Gamma_{22}^{(i)} T
+ \alpha DT^2
\label{g121}\\
\Gamma_{22}^{(f)} &=&  \Gamma_{22}^{(i)} + 2DT \,.
\label{g221}
\eea
The $T^3$ contribution to the variances has also been discussed
by Rhines and Young (1983) in the context of fluid mixing.

For the reversal we allow for a different shear rate $\alpha'$
and some perturbation in the velocity field. If ${\bf q}_f$
is an initial displacement in the trajectory and $\delta{\bf u}
= (u_1, u_2)$ a constant perturbation to the velocity field,
the displacement of the echo will be
\bea
q_{1,e}&=& q_{1,f} - (u_1-\alpha' q_{2,f})t - \alpha' u_2 t^2/2\\
q_{2,e}&=& q_{2,f} - \alpha' u_2 t - \alpha' u_2 t^2/2\,.
\eea
The variances become
\bea
\Gamma_{11}^{(e)} &=&  \Gamma_{11}^{(i)} 
	+ 2(\alpha-\alpha') \Gamma_{12}^{(i)} T 
	+  (\alpha-\alpha')^2 \Gamma_{22}^{(i)} T^2 \nonumber\\
&\ &  + 4DT + 
	\frac{1}{3}(2\alpha^2+8\alpha'^2-6\alpha\alpha') DT^3
\label{g112}\\
\Gamma_{12}^{(e)} &=&  \Gamma_{12}^{(i)} + 
(\alpha-\alpha') \Gamma_{22}^{(i)} T + (\alpha-3\alpha') DT^2
\label{g122}\\
\Gamma_{22}^{(e)} &=&  \Gamma_{22}^{(i)} + 4DT \,.
\label{g222}
\eea
This result for the variances may be verified for a few limiting cases:
(i) Without shear $\alpha=\alpha'=0$ the diagonal elements
increase like $4DT$ as for regular diffusion over a time interval $2T$.
(ii) Without diffusion (D=0) the determinant of the matrix does not 
change.
(iii) Without diffusion (D=0) and equal shear in the forward and backward 
direction ($\alpha=\alpha'$) the reversal is perfect and the initial
variances are restored.
(iv) The parameters $\alpha'=-\alpha$ correspond to the  
situation that the backward integration is just a 
continuation of the forward integration and the 
expressions (\ref{g112})-(\ref{g222}) agree with 
(\ref{g111})-(\ref{g221}) for an evolution time of $2T$.
(iv) For equal shear $\alpha=\alpha'$ in forward and 
backward direction the variances are
\bea
\Gamma_{11}^{(e)} &=&  \Gamma_{11}^{(i)} + 4DT + 
\frac{4}{3}\alpha^2 DT^3\\
\Gamma_{12}^{(e)} &=&  \Gamma_{12}^{(i)} - 2\alpha DT^2\\
\Gamma_{22}^{(e)} &=&  \Gamma_{22}^{(i)} + 4DT \,;
\eea
the prefactor of the cubic term in $\Gamma_{11}^{(e)}$ is smaller than
would be obtained from (\ref{g112}) for a time $2T$, 
indicating that the reversal of the shear induced
broadening is partial and not complete.

The different time regimes in the variances are easily identified.
Consider the terms linear in time first: diffusion will be
noticable on a time scale
$T_D\approx 1/D$, the differences in 
the shear rates on a time scale $T_\delta \approx 1/|\alpha-\alpha'|$.
Thus, a large diffusion can swamp the effects from the difference
between the two Hamiltonians. For the nonlinear terms, the one with
the difference in shear rates appears around $T_\delta$, and
the one with the cubic term in diffusion near $1/\alpha$. In typical
applications the latter should be smaller than $T_\delta$.

The classical fidelity contains the determinants of the variances.
In the absence of diffusion, $D=0$, we have
$\det \Gamma_e = \det \Gamma_i$ and 
\beq
\det (\Gamma_i + \Gamma_e) = 4 \det \Gamma_i + (\alpha-\alpha')^2 T^2
\Gamma_{22}^{(i)}
\eeq
so that the fidelity has a $1- \mbox{const}\cdot T^2$ behaviour for 
short times and a $1/ (|\alpha-\alpha'|T)$ 
decay for times longer than $T_\delta$. With diffusion and
equal forward and backward shear, $\alpha=\alpha'$, we have
\bea
\det \Gamma_e &=& \det \Gamma_i + 4 (\Gamma_{11}^{(i)}+\Gamma_{22}^{(i)}) DT \cr
& & 
+ 16 D^2 T^2 + 4 \Gamma_{12}^{(i)} \alpha D T^2 + \frac{4}{3}
\Gamma_{12}^{(i)} \alpha^2 D T^3 + \frac{4}{3} \alpha^2 D^2 T^4
\eea
and
\bea
\det \left( \Gamma_i+\Gamma_e \right) &=& 4 \det \Gamma_i + 8 (\Gamma_{11}^{(i)}+\Gamma_{22}^{(i)}) DT \cr
& & 
+ 16 D^2 T^2 + 8 \Gamma_{12}^{(i)} \alpha D T^2 + \frac{8}{3}
\Gamma_{12}^{(i)} \alpha^2 D T^3 + \frac{4}{3} \alpha^2 D^2 T^4
\eea
so that a short-time behaviour $1- \mbox{const} \cdot t$ 
and a long-time behaviour of $1/(\sqrt{\alpha D} T)$ follow.

The contributions from displacements enter in the exponent and
can in principle introduce rapid decays. Consider, e.g.,
the case $\alpha=\alpha'$ and weak diffusion, weak shear and 
short times, so that $DT$ and $\alpha T$ are 
smaller than the initial variances. Then
$\Gamma^{(e)}\approx \Gamma^{(i)}$ and the only time dependence
will come from the perturbations ${\bf q}_0$ and $\delta u$:
the exponent will contain a polynominal of fourth order
in time, and this leads to a rapid and faster than
exponential decay. If diffusion is added the increase in variance 
can compensate part of the displacement growth, but for strong
diffusion and large times an exponential decay will remain:
the square of ${\bf q}^2$ will increase like $T^4$ and
the variance increases like $T^3$, so that the ratio
increases linearly, giving an exponential decay.

Thus, without displacement the overlap integral decays algebraically
as contained in the prefactor, but with displacement the decrease 
can be dramatic when the densities are separated by
more than their widths.

\subsection{Chaotic systems}
As a model for a chaotic system take a simple
hyperbolic motion, ${\bf u} = (\lambda x, - \lambda y)$
and assume an initial density aligned with the
unstable ($x$-) and stable ($y$-) direction, i.e.
$\Gamma_{12}^{(i)}=0$. After forward evolution we have
\bea
\Gamma_{11}^{(f)} &=&  \Gamma_{11}^{(i)} e^{2\lambda T} 
+ \frac{D}{\lambda} (e^{2\lambda t}-1)
\label{exp1}\\
\Gamma_{22}^{(f)} &=&  \Gamma_{22}^{(i)} e^{-2\lambda T} 
+ \frac{D}{\lambda} (1-e^{-2\lambda t}) \,.
\label{exp2}
\eea
Thus, there is an exponential contraction down to the
limit set by diffusion.
If the backward integration has a slightly different 
stretching rate $\lambda'$, then 
for the echo
\bea
\Gamma_{11}^{(e)} &=&  \Gamma_{11}^{(i)} e^{2(\lambda-\lambda') T} 
+ \frac{D}{\lambda} (e^{2\lambda T}-1) e^{-2\lambda'T}
+ \frac{D}{\lambda'} (1-e^{-2\lambda' T})\\
\Gamma_{22}^{(e)} &=&  \Gamma_{22}^{(i)} e^{-2(\lambda-\lambda') T} 
+ \frac{D}{\lambda} (1-e^{-2\lambda T}) e^{2\lambda' T}
+ \frac{D}{\lambda'} (e^{2\lambda' T}-1) \,.
\eea
As in the previous case we can study various limiting situations,
such as $\lambda,\lambda'\rightarrow 0$, where linear diffusion
results, or $\lambda=-\lambda'$, where the expressions (\ref{exp1})
and (\ref{exp2})
for times up to $2T$ are recovered. 
When forward and backward stretching
rates are the same, $\lambda=\lambda'$, 
the variances become
\bea
\Gamma_{11}^{(e)} &=&  \Gamma_{11}^{(i)} 
+ 2\frac{D}{\lambda} (1-e^{-2\lambda T}) \\
\Gamma_{22}^{(e)} &=&  \Gamma_{22}^{(i)}
+ 2\frac{D}{\lambda'} (e^{2\lambda' T}-1) \,.
\eea
Note that the variance in $x$ has hardly changed
whereas the one in $y$ grows exponentially. 
The reason is that the $x$-variance
grows during the forward integration and collapses then during
the backward evolution, down to a limit set by the diffusional
broadening. The exponential growth and contraction is thus almost
perfectly compensated. For the $y$-variance we have first the 
contraction, down to the limit set by diffusion. The expansion
during the backward evolution then starts from this finite
amplitude, and not from the exponentially small contraction
of the deterministic evolution of the initial variance. 
As a result, the echo is stretched out along the direction
that was the stable one during forward evolution. If forward
and backward evolution are interchanged, then so is
the orientation of the spreading of the density: it will then
point in the $x$-direction.
For this growth to be noticable the time evolution has 
to be followed for times longer than about 
$(\ln(\lambda \Gamma^{(i)}/D))/(2\lambda)$. 

The behaviour of the classical fidelity for short times
is quadratic or linear, for $D=0$ and $D\ne0$, respectively.
On longer times there is an exponential decay, like
$\exp(-|\lambda-\lambda'|T)$ without diffusion and like
$\exp(-\lambda'T)$ with diffusion.

The appearance of differences between Lyapunov exponents
reflects a relation between forward and backward flow:
the stable and unstable manifolds are aligned. A more general
difference between forward and backward flow will
break this alignment and introduce exponentials in
$\lambda'$.

Small perturbations in position and in the velocity fields
will grow exponentially. Eq.~(\ref{displace}) gives
\bea
q_{1,e}&=& (q_{1,f}-u_1/\lambda')e^{-\lambda' T} + u_1/\lambda'\\
q_{2,e}&=& (q_{2,f}+u_2/\lambda')e^{\lambda' T} - u_2/\lambda' \,.
\eea
The contributions from the displacement to the decay are weaker
than in the linear shear flow, since the stretching of the
variances is in the same directions as the separation of 
trajectories, so that an exponential increase in $q_2$ can be
compensated by an exponential increase in variances.
However, in the absence of diffusion and for the 
same Lyapunov exponents in forward and backward direction
an exponentially growing displacement can lead to 
a drastic drop off, like $\exp(-\exp \lambda t)$, in overlap, 
simply because the 
Gaussians are shifted relative to each other.

\section{Final remarks}
Already the simple examples in the previous section
show a wide range of dynamical behaviour in 
classical echo experiments. There are two ingredients:
the variation in variances and a displacement
between initial and echo density. Without displacement 
the decay in the integrable system is slower than in 
the chaotic one, and the same applies when there is 
a displacement in both systems. However, the overlap
can drop off faster in an integrable system with displacement 
compared to a chaotic one without. 

The drop off from the displacement is connected with the 
Gaussian shape of the densities: if the tails fall off
more slowly then also the overlap will decay more slowly
as a function of displacement. The shape dependence
should be stronger in integrable systems than in chaotic
ones, since the stretching of variances and shapes
goes in parallel with the exponential growth of 
displacement.

The calculations are based on Gaussian densities and 
linearizations of the flow fields near trajectories. This becomes
questionable if the densities spread out too far, a problem
that occurs more likely and more quickly in chaotic systems
than in integrable systems. In integrable systems the
dangerous terms come from quadratic dependencies of the 
winding frequencies on action and from curvatures introduced when
mapping the tori onto position space. The density will then
coil up in whirls. In a 2-d of freedom system
this can be expected to happen linearly in time. In chaotic
systems the linear approximation is applicable if the 
stable and unstable manifolds are close to straight lines within
the area covered by the density and if the variations in 
stretching and contraction rate are small. Typically,
this will limit the time interval to a logarithmically
short one. For longer times the density develops tendrils
that follow the manifolds as they wiggle through phase space
(for an early discussion of such effects within the quantum
maps, see Berry et al, 1979). In both cases the degree
to which the Gaussian approximation breaks down depends
on details of the nonlinear contributions and has to be
considered for specific models.

The transformation to a comoving frame eliminates the 
center of mass motion and emphasizes the linearized dynamics
near the trajectory. The same discussion thus applies
to stationary points. Two examples, shear flow or
parabolic type and hyperbolic type, have been discussed
here, the third class, elliptic type, has periodically
oscillating variances.

Two-dimensional flows provide an ideal testing ground for the
results presented here. Localized spots of dye can be prepared as 
initial densities and their motion can be followed in various
2-d flows (Aref 1984, 2002, Ottino 1990, Homsey 2001, Chaiken 1986).
Elliptic and hyperbolic points can be realized
most easily in cellular flows (J\"utner \etal 1997, Williams \etal 1997,
Rothstein \etal 1999) 
Molecular diffusivities
are fixed by the selection of dye and solvent,
but variations of shear rate through the amplitude of the 
velocity fields and initial variances through the size of the
spot provide enough degrees of freedom to explore the 
full range in behaviour.  In particular the 2-d Lorentz force
driven flows (J\"utner \etal 1997, Williams \etal 1997,
Rothstein \etal 1999)  should have enough 
flexibility to study echoes in flows and to provide 
quantitative tests of the various expressions derived here.
It should be possible to see the cubic laws in the variances
in integrable systems, the chaos assisted spreading in chaotic
systems and the dependence on the order of forward/backward
propagation. 

It would also be of interest to reconsider the experimental
protocol of Chaiken \etal (1986): they followed a closed line of dye.
After reversal the line was displaced left and right of the 
original trace and also had slightly different widths. This suggests
different histories of the center of mass of small line elements
and passage through
regions with different degrees of chaos. Using the complete
tracings of the flow field as in (Voth, Haller and Gollub 2002) it should be 
possible to pin down the different regions and interactions
and to characterize the reversal completely.

\section*{Acknowledgment}
I would like to thank Th. Seligman and Th. Gorin for 
stimulating discussions on the relation between classical 
and quantal fidelity and their symmetries.



\section*{References}
\begin{thebibliography}{00}

\bibitem{Aref84} 
Aref H 1984, J. Fluid. Mech. {\bf 143}, 1 

\bibitem{Aref02} 
Aref H 2002, Phys. Fluids {\bf 14}, 1315--1325 

\bibitem{Berry}
Berry MV, Balazs NL, Tabor M and Voros A 1979, 
Ann. Phys (NY) {\bf 122}, 26--63

\bibitem{Casati02}
Benenti G and Casati G 2002,
Phys. Rev. E {\bf 65}, 066205 (4 pages)

\bibitem{Purcell54}
Carr HY and Purcell EM 1954, 
Phys. Rev. {\bf 94} 630

\bibitem{Casati86}
Casati G, Chirikov BV, Guarneri I and Shepelyansky DL 1986, 
Phys. Rev. Lett. {\bf 80} 2437--2440

\bibitem{Tomsovic}
Cerrutti NR and Tomsovic S 2002,
Phys. Rev. Lett. {\bf 88}, 054103 (4 pages)

\bibitem{Chaiken}
Chaiken J, Chevray R, Tabor M and Tan OM 1986 
Proc. R. Soc. (London) A{\bf 408}, 165

\bibitem{Eckhardt90}
Eckhardt B 1990, in 'Topological Fluid Mechanics', HK Moffat and
A Tsinober (eds), Cambridge University Press, Cambridge, p. 23-33

\bibitem{Hahn50}
Hahn EL 1950, Phys. Rev. {\bf 80} 580

\bibitem{Homsey00}
Homsey GM \etal 2000,
	Multimedia Fluid Mechanics CD, Cambridge University Press,
	Cambridge 

\bibitem{Jacquod}
Jacquod P, Silvestrov PG and Beenakker CWJ 2001,
Phys. Rev. E{\bf 64} 055203 (R)

\bibitem{Pastawski01}
Jalabert R and Pastawski HM 2001,
Phys. Rev. Lett. {\bf 86} 2490--2493

\bibitem{Juetner}
J\"utner B, Marteau D, Tabeling P and Thess A, 1997,
Phys. Rev. E {\bf 55} 5479

\bibitem{Pastawski95a}
Levstein PR, Usaj G and Pastawski HM 1998, 
J. Chem. Phys. {\bf 108} 2718--2724

\bibitem{Loschmidt}
Loschmidt J 1876, Sitzungsber. Kais. Akad. Wiss. Wien Math.
Naturwiss. Klasse {\bf 73} 128--142

\bibitem{Thomas}
Niggemeier W, von Plessen G, Sauter S and Thomas P 1993,
Phys. Rev. Lett. {\bf 71}, 770-772

\bibitem{Ottino}
Ottino JM 1990, 
Annu. Rev. Fluid. Mech. {\bf 22}, 207

\bibitem{Pastawski95}
Pastawski GM, Levstein PR and Usaj G 1995, 
Phys. Rev. Lett. {\bf 75} 4310--4313

\bibitem{Peres84} 
Peres A 1984, Phys. Rev. A{\bf 30}, 1610--15

\bibitem{Prosen1}
Prosen T and Znidaric M 2002, 
J. Phys. A {\bf 35}, 1455--1481

\bibitem{Prosen2}
Prosen T and Seligman T 2002,
J. Phys. A {\bf 35}, 4707--4727

\bibitem{Rhines}
Rhines PB and Young WR 1983, J. Fluid Mech. {\bf 133}, 133--145

\bibitem{Rothstein}
Rothstein D, Henry E and Gollub JP, 1999
Nature {\bf 401} 770


\bibitem{Voth}
Voth GA, Haller G, and Gollub JP 2002,
Phys. Rev. Lett. {\bf 88}, 254501 (4 pages)

\bibitem{Williams}
Williams BS, Marteau D and Gollub JP 1997,
Phys. Fluids {\bf 9}, 2061

\bibitem{Wisniacki}
Wisniacki DA and Cohen D 2001,
preprint $nlin.CD/0111125$




\end{thebibliography}
\end{document}